\PassOptionsToPackage{compatibility=false,justification=centering,font=normalsize,labelfont=sf,textfont=sf}{caption}
\pdfoutput=1
\documentclass[journal]{IEEEtran}
\usepackage{amsmath,amsfonts}
\usepackage{algorithm}
\usepackage{array}
\usepackage[caption=false,font=normalsize,labelfont=sf,textfont=sf]{subfig} 
\usepackage{textcomp}
\usepackage{stfloats}
\usepackage{url}
\usepackage{multirow}
\usepackage[table,xcdraw]{xcolor}
\usepackage{booktabs}
\usepackage{verbatim}
\usepackage{graphicx}
\usepackage{gbt7714}
\usepackage{authblk}
\begin{document}

\title{VocalCrypt: Novel Active Defense Against Deepfake Voice Based on Masking Effect}

\author{
    Qingyuan Fei\thanks{Qingyuan Fei, Wenjie Hou, Xuan Hai and Xin Liu are with the School of Information Science and Engineering, Lanzhou University, Lanzhou 730000, Gansu, China.} \thanks{Qingyuan Fei and Wenjie Hou contributed equally to this work.}
    Wenjie Hou
    Xuan Hai
    Xin Liu$^{*}$\thanks{Corresponding Authors: Xin Liu (email: bird@lzu.edu.cn)}
}



\maketitle

\begin{abstract}
The rapid advancements in AI voice cloning, fueled by machine learning, have significantly impacted text-to-speech (TTS) and voice conversion (VC) fields. While these developments have led to notable progress, they have also raised concerns about the misuse of AI VC technology, causing economic losses and negative public perceptions. To address this challenge, this study focuses on creating active defense mechanisms against AI VC systems.

We propose a novel active defense method, VocalCrypt, which embeds pseudo-timbre (jamming information) based on SFS into audio segments that are imperceptible to the human ear, thereby forming systematic fragments to prevent voice cloning. This approach protects the voice without compromising its quality. In comparison to existing methods, such as adversarial noise incorporation, VocalCrypt significantly enhances robustness and real-time performance, achieving a 500\% increase in generation speed while maintaining interference effectiveness.

Unlike audio watermarking techniques, which focus on post-detection, our method offers preemptive defense, reducing implementation costs and enhancing feasibility. Extensive experiments using the Zhvoice and VCTK Corpus datasets show that our AI-cloned speech defense system performs excellently in automatic speaker verification (ASV) tests while preserving the integrity of the protected audio.

\end{abstract}

\begin{IEEEkeywords}
AI clone speech, active defense, sound copyright protection, voice counterfeit defense, and voice security.

\end{IEEEkeywords}

\section{Introduction}
\label{sec:intro}
In recent years, advancements in deep learning technology have led to remarkable progress in audio processing, particularly in areas such as automatic speaker verification (ASV), automatic speech recognition (ASR), and speaker identification \cite{morris2017music, hagen2015playlist, datta2018spotify, wlomert2019streaming}. The ongoing advancement of these technologies has greatly enhanced the performance and expanded the scope of applications in audio processing. However, this rapid development also introduces potential risks of misuse, with the emergence of Deepfake technology being a particularly pressing concern \cite{sahidullah_comparison_2015, dolhansky_deepfake_2020, li_exposing_2019, ciftci_fakecatcher_2020}. Deepfake speech poses a serious threat to personal privacy, property safety, and social reputation \cite{noauthor_artificial_nodate, chou2019one}. Voice conversion (VC) \cite{qian2019autovc} and Text-to-Speech (TTS) \cite{shen2018natural, ren2019fastspeech} technology are important tools for generating Deepfake speech forgeries. Existing Deepfake forgery detection methods, such as sound watermarking techniques \cite{zhang2023speech, wang2023robust, chen2024deepmark} and passive detection technology \cite{ahmed_void_2020, gao_audio_2010, lieto_hello_2019, wang_deepsonar_2020}, can identify counterfeit content to some extent. However, existing methods that rely on passive detection techniques \cite{zhang2024deepsonar, wang2023detecting, chen2023voiceprint, monteiro2023real, liu2023survey} have demonstrated effectiveness in identifying Deepfake speech after it has been disseminated. These methods, although useful in post-attack detection, function mainly as remedial measures. However, they exhibit significant limitations: (1) they provide solutions only after the attack has occurred, which makes it difficult to prevent immediate harm; and (2) they are not tailored to detect specific attack features, often overfitting to certain characteristics of previously observed attack techniques, which severely restricts their generalizability. As a result, they often fail to identify unknown attacks because they rely heavily on the identification of known attack signatures. These detection methods are only effective against attacks with known signatures, making them inadequate for detecting unknown or novel attacks. To address the above issues, this paper introduces an innovative approach that integrates both offensive and defensive strategies by targeting vulnerabilities in AI VC systems. The proposed method, VocalCrypt, introduces pseudo-tones in the inaudible regions of the audio to disrupt cloned speech. This approach leverages the masking effect to prevent the AI VC system from learning the true characteristics of the target voice, instead misguiding it with the pseudo-tones. In addition to its defensive nature, VocalCrypt focuses on early detection and the identification of specific attack features, improving the system's generalizability and minimizing harm.

While existing methods, such as those based on adversarial perturbations and GANs, have demonstrated effectiveness in adding defense layers to mel spectrograms, they have notable limitations. These solutions are often vulnerable to attacks like compression, noise reduction, and ablation, and they lack real-time applicability, limiting their use in large-scale, real-world scenarios. Furthermore, they typically target specific AI VC systems, lacking broad generalizability. In contrast, VocalCrypt's use of pseudo-tones not only offers a more robust and scalable defense against a wider range of AI VC models but also resists noise reduction, providing a more comprehensive and practical solution for real-time protection against attacks where attackers attempt to remove defensive perturbations through audio denoising techniques, thereby improving the overall defense effectiveness and system robustness. Inspired by Moore \cite{moore2003factors} and the research of others, these studies have shown that in typical complex sound scenes (such as music or speech), about 30\% to 60\% of the audio signal can be masked and imperceptible to the human auditory system \cite{moore2003factors, bosi1997iso, painter1996perceptual}. Based on this, we have designed a pseudo-tone embedding framework based on the masking effect, which significantly improves robustness and generation speed while ensuring defense effectiveness. In the experiment, we used the CSTR and VCTK Corpus \cite{veaux_cstr_2017} two public datasets that were extensively tested to verify the effectiveness of the proposed method.

In summary, the main contributions of this paper include the following three aspects:
\begin{enumerate}
\item We analyzed existing AI VC detection technologies and identified key limitations, including reliance on post-detection, lack of real-time capabilities, and limited robustness. To address these, we proposed a proactive defense strategy that transforms passive detection into intervention by exploiting vulnerabilities in AI VC systems, offering a more effective and robust solution.
\item We introduced VocalCrypt, a universal active defense solution for AI VC—the first of its kind in this field. Compared to traditional methods, it offers superior flexibility and enables real-time active defense against AI-cloned voices.
\item We tested our pseudo-voice interference method on five advanced models using Chinese and English datasets. Our method outperformed existing approaches in benchmark tests, showing significant advantages against noise and compression attacks, with computational efficiency 500\% higher.
\end{enumerate}

\section{Background}
\subsection{Deepfake Voice}
Voice cloning aims to create a synthetic voice that closely resembles the tone, timbre, and characteristics of a target individual's voice. This process mainly depends on two key techniques: voice conversion and TTS generation. Voice conversion works by adjusting the speech signal of an arbitrary speaker to match that of the target speaker while maintaining the original linguistic content. In contrast, TTS is more flexible, capable of generating the desired speech for any given text without the need for the original speaker's voice.
\subsubsection{Text Processing}

For TTS systems to produce high - quality speech, the input text needs to go through a series of preprocessing steps to extract both linguistic and acoustic features. The preprocessing includes text normalization and linguistic analysis. Text normalization aims to standardize the text to make it suitable for further processing by the TTS system. This involves operations such as removing punctuation, converting numbers into spoken forms, and expanding abbreviations. The subsequent linguistic analysis step is to segment the normalized text into basic linguistic units such as sentences, phrases, and words. These units are then converted into phonemes, the basic sound units of a language, and tokenized for model input. Finally, the tokenized phonemes are used to generate the synthesized speech. Advanced TTS models often incorporate prosody prediction, which determines the rhythm, stress, and intonation patterns to make the output more natural. Additionally, some systems integrate syntactic and semantic analysis to help the TTS model better understand the text structure and meaning for more accurate speech synthesis \cite{Mansfield2019}.
\subsubsection{Acoustic Model}

The acoustic model is a crucial component in TTS systems. It converts linguistic data into acoustic features, such as spectrograms, which define the sound of the synthesized speech. Deep neural network - based acoustic models have significantly outperformed traditional statistical models. Wang et al. \cite{Wang2017} introduced Tacotron, the first deep - learning - based model for end - to - end speech synthesis. Tacotron uses recurrent neural networks (RNNs) to simulate the dynamic nature of speech signals and employs an attention mechanism to align the input and output. However, Tacotron suffers from some distortion and noise. To address this, Shen et al. \cite{Shen2018} proposed Tacotron 2, which incorporates a position - sensitive attention module, greatly improving the synthesis quality. Although both Tacotron and Tacotron 2 can produce high - quality results, they are computationally intensive. To solve this efficiency problem, Ping et al. \cite{Ping2018} introduced FastSpeech, a model that uses a fully - convolutional sequence - to - sequence architecture, enabling parallel generation of mel - spectrograms. FastSpeech 2 \cite{Ren2020} further improves on this by incorporating an acoustic prior, enhancing the model's output. In relevant research, Tacotron 2 \cite{Shen2018} and FastSpeech 2 \cite{Ren2020} are often used as the default acoustic models, but the solution can also be adapted to other models.
\subsubsection{Vocoder}

Once the spectrograms are generated, a vocoder synthesizes the final speech signal based on frequency band and pitch information. The Griffin - Lim algorithm \cite{Griffin1984} is a commonly used vocoder that reconstructs the speech signal from the Short - Time Fourier Transform (STFT). Although it is computationally efficient, the Griffin - Lim algorithm may introduce obvious background noise and distortion, and its performance is highly sensitive to parameter selection, which may require fine - tuning. To improve the quality of the generated audio, deep - learning - based vocoders such as WaveGAN \cite{Yamamoto2020} and HiFi - GAN \cite{Kong2020} have been developed. WaveGAN uses convolutional and deconvolutional layers to generate high - quality audio signals, but sometimes it may produce unstable or distorted outputs, especially for complex or long signals. In contrast, HiFi - GAN uses an advanced training process, including feature - matching loss functions and a multi - scale discriminative model, enabling it to generate more natural and accurate audio than earlier GAN - based methods. In relevant work, the Griffin - Lim algorithm \cite{Griffin1984} and HiFi - GAN \cite{Kong2020} are often used as the default vocoders. In addition, VITS \cite{Kim2021}, which takes text as input and directly generates the speech output in an end - to - end manner, is also considered.
Active Defense Based on Acoustic Masking Effect
Previous research has mainly focused on voice cloning technology itself. In contrast, we take a different approach and propose an active defense method against AI - cloned voices based on the acoustic masking effect. The acoustic masking effect is an important phenomenon in psychoacoustics, where a stronger sound can make a weaker sound imperceptible. We innovatively apply this effect to defend against AI - cloned voice attacks. Specifically, the defense system, through careful design, uses the principle of acoustic masking to interfere with and resist potential cloned voice attacks. Without affecting the intelligibility of normal speech, the voice signal that may be subject to cloning attacks is specially processed so that when it is cloned, it generates interference that is difficult to eliminate, effectively reducing the quality and credibility of the cloned voice and achieving active defense. The following sections will elaborate on the specific implementation, experimental verification, and comparison analysis with existing defense methods of this method.
\subsection{Masking  Effect}
The masking effect, a crucial phenomenon in psychoacoustics, occurs when a stronger sound renders a weaker sound imperceptible. This phenomenon manifests in two primary forms: frequency masking and temporal masking. Frequency masking refers to the spectral-domain suppression of weak sounds by stronger signals with similar frequency components \cite{zwicker1990psychoacoustics}. Temporal masking encompasses both pre-masking (forward temporal masking) and post-masking (backward temporal masking), describing how dominant sounds can obscure weaker signals occurring either before or after their presence in the temporal domain \cite{bosi2002introduction,brandenburg1999mp3}.

Early masking threshold models were established through foundational psychoacoustic research by Zwicker \cite{zwicker1990psychoacoustics}, while Johnston's subsequent work developed a perceptually transparent encoding algorithm that operationalized these principles for practical audio compression systems \cite{johnston1988transform}. Johnston's algorithm specifically demonstrated how psychoacoustic models could be implemented to achieve significant improvements in both coding efficiency and perceptual audio quality \cite{johnston1988transform}. Further research by Bosi and Goldberg quantified the residual effects of temporal masking, particularly analyzing post-masking persistence durations, which are critical for frame-length optimization in audio codecs \cite{bosi2002introduction,brandenburg1999mp3}. The residual effect of masking after the cessation of a sound has been thoroughly examined, underscoring its significance for frame processing in audio coding.

Masking effects are widely leveraged in audio compression, particularly in the MPEG standard, where psychoacoustic principles are employed to eliminate imperceptible audio data. This approach achieves efficient compression while maintaining high perceptual audio quality. Studies have shown that between 30\% and 70\% of audio signals fall below the masking threshold, which justifies the rationale for our approach of incorporating pseudo-timbres based on the masking threshold.

Therefore, we are extending this method to active defense mechanisms.
\section{Motivation and Insight}

In the field of speech conversion defense, Huang\cite{huang_defending_2021}
et al. proposed an innovative method to protect personal speech identity through adversarial attacks. They designed three different adversarial attack strategies, namely end-to-end attacks, embedding attacks, and feedback attacks, to address the privacy risks that may be caused by speech conversion technology. These methods effectively destroy the authenticity of the fake speech by adding tiny adversarial perturbations to the audio that are imperceptible to humans, making it impossible for the speech conversion model to accurately imitate the target speaker.
In addition, Dong et al\cite{dong_active_2024}. proposed a framework based on generative adversarial networks (GANs) that aims to quickly generate adversarial perturbations to defend speech conversion models. Their method enhances the effectiveness of the defense by adding adversarial perturbations to the mel spectrogram of the target audio and designing a spectral waveform conversion simulation module to simulate the reconstruction process from the adversarial mel spectrogram to the waveform. Compared with the traditional iterative method, the GAN-based solution excels in both generation speed and defense performance, and has achieved good defense results on multiple speech conversion models in both white-box and black-box scenarios.
However, there are three serious problems with the GAN-based active defense solution: 1. Poor robustness, vulnerable to compression attacks, noise reduction attacks, and ablation attacks. 2. Not real-time, cannot be used on a large scale in real-world scenarios. 3. Poor generality, only performs adversarial attacks against a certain AI clone speech system, and is not widely applicable. This led to the proposed method of adding pseudo-tone based on masking thresholds.

\section{Our Approach}
\subsection{Overview}
To safeguard against voice conversion (VC) and text-to-speech (TTS) attacks, we propose an active defense mechanism based on pseudo-timbre embedding that leverages the auditory masking effect. First, both the original audio and the pseudo-timbre signals are preprocessed by applying a Discrete Wavelet Transform (DWT) to obtain their localized time-frequency representations. Next, to fully exploit the masking effect—as detailed in the masking threshold calculation module in the upper part of Fig. 1—the original audio is decomposed into 25 frequency bands via a filter bank, and for each critical band, the absolute masking threshold is computed using the spectral flatness measure (SFM). Since the majority of the audio energy lies in the low-frequency region and the human ear is less sensitive to these frequencies, only the information in critical bands 1 through 7 is utilized for embedding. The corresponding pseudo-timbre components for these bands are then embedded into the original audio using Quantization Index Modulation (QIM). Finally, the protected audio is reconstructed by applying the Inverse Discrete Wavelet Transform (IDWT) on the modified signal. This method not only offers robust protection against VC and TTS attacks but also preserves the perceptual quality of the original audio.
\begin{figure*}[htbp]  
  \centering
  \includegraphics[scale=0.6]{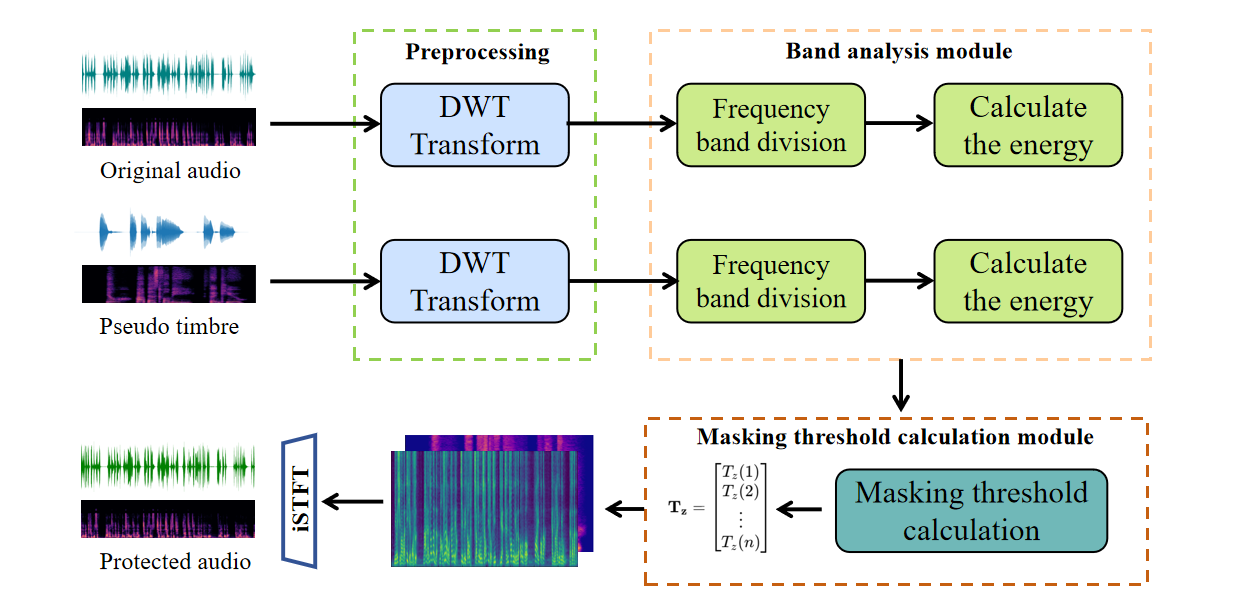}  
  \captionsetup{justification=centering}  
  \caption{Architecture of the VocalCrypt}
  \label{fig:image}
\end{figure*}
\subsection{Critical Band Division}
In this study, the need for dividing the critical bands arises from the nonlinear frequency response characteristics of the human auditory system. The human ear exhibits significant sensitivity differences to sounds of varying frequencies, especially with regard to frequency resolution and the masking effect, both of which display critical band dependence \cite{zwicker1961subdivision}. For example, auditory resolution is higher in the low-frequency range, whereas it gradually decreases with increasing frequency. This characteristic makes traditional linear frequency divisions inadequate for accurately describing auditory perception mechanisms. To maximize the pseudo-tonal embedding effect, this paper adopts a critical band division method based on the Bark scale, segmenting the auditory range from 20 Hz to 22.05 kHz into 25 bands (as shown in Table \ref{tab:masking_threshold}). The basis for this division lies in the fact that the Bark scale, by defining the "critical bandwidth," transforms frequency into a psychoacoustic unit that approximates auditory resolution \cite{zwicker1961subdivision}. Specifically, when the frequency separation between two components is smaller than a certain critical bandwidth, the human ear finds it difficult to distinguish them as separate entities and perceives them as a combined signal within the same band. Therefore, the Bark-based division not only aligns with the nonlinear characteristics of auditory perception but also enhances the efficiency of signal processing algorithms.
\begin{table}[htbp]
\caption{The Absolute Masking Threshold}
\begin{center}
\begin{tabular}{cccc}
\toprule
\multirow{2}{*}{\textbf{Bark Zone}} & \multicolumn{3}{c}{\textbf{Frequency (Hz)}} \\ 
\cmidrule(l){2-4}
 & \textbf{Center} & \textbf{Low End} & \textbf{High End} \\ 
\midrule
1 & 50  & 20  & 100 \\
2 & 150 & 100 & 200 \\
3 & 250 & 200 & 300 \\
4 & 350 & 300 & 400 \\
5 & 450 & 400 & 510 \\
6 & 570 & 510 & 630 \\
7 & 700 & 630 & 770 \\ 
\bottomrule
\end{tabular}
\label{tab:masking_threshold}
\end{center}
\end{table}
Energy is primarily concentrated in the low-frequency bands, and the human ear exhibits reduced sensitivity to low-frequency signals. Consequently, only the critical frequency bands corresponding to $1 \sim 9$ are selected for embedding information, while the other critical frequency bands are disregarded. The frame length during dynamic segmentation is denoted as $L$. Each frame signal is initially transformed using a 3-layer Haar wavelet, with the corresponding frequency bands of the 3-layer scale coefficients $c_3(i)$ and the 3-layer wavelet coefficients $d_3(i)$ being $0 \sim 500 \ \text{Hz}$ and $500 \sim 1000 \ \text{Hz}$, respectively. A Discrete Cosine Transform (DCT) is then applied to both $c_3(i)$ and $d_3(i)$, yielding $L/8$ DCT coefficients, denoted as $x_c(k)$ and $x_d(k)$, respectively. These DCT coefficients are equally spaced to divide the low-frequency bandwidth, with a frequency resolution of $\frac{f_s}{2L}$ (Hz). The frequency points represented by $x_c(k)$ and $x_d(k)$ are then allocated to each Bark domain, thereby achieving the critical band division.

In this paper, we set $L = 504$, which results in a frequency resolution of approximately $7.94 \ \text{Hz}$. By referencing the upper and lower boundaries of each critical frequency band as given in Table 1, we can determine the frequency range and the number of DCT coefficients contained within each critical frequency band.
\begin{figure}[htbp]  
  \centering
  \includegraphics[width=\linewidth]{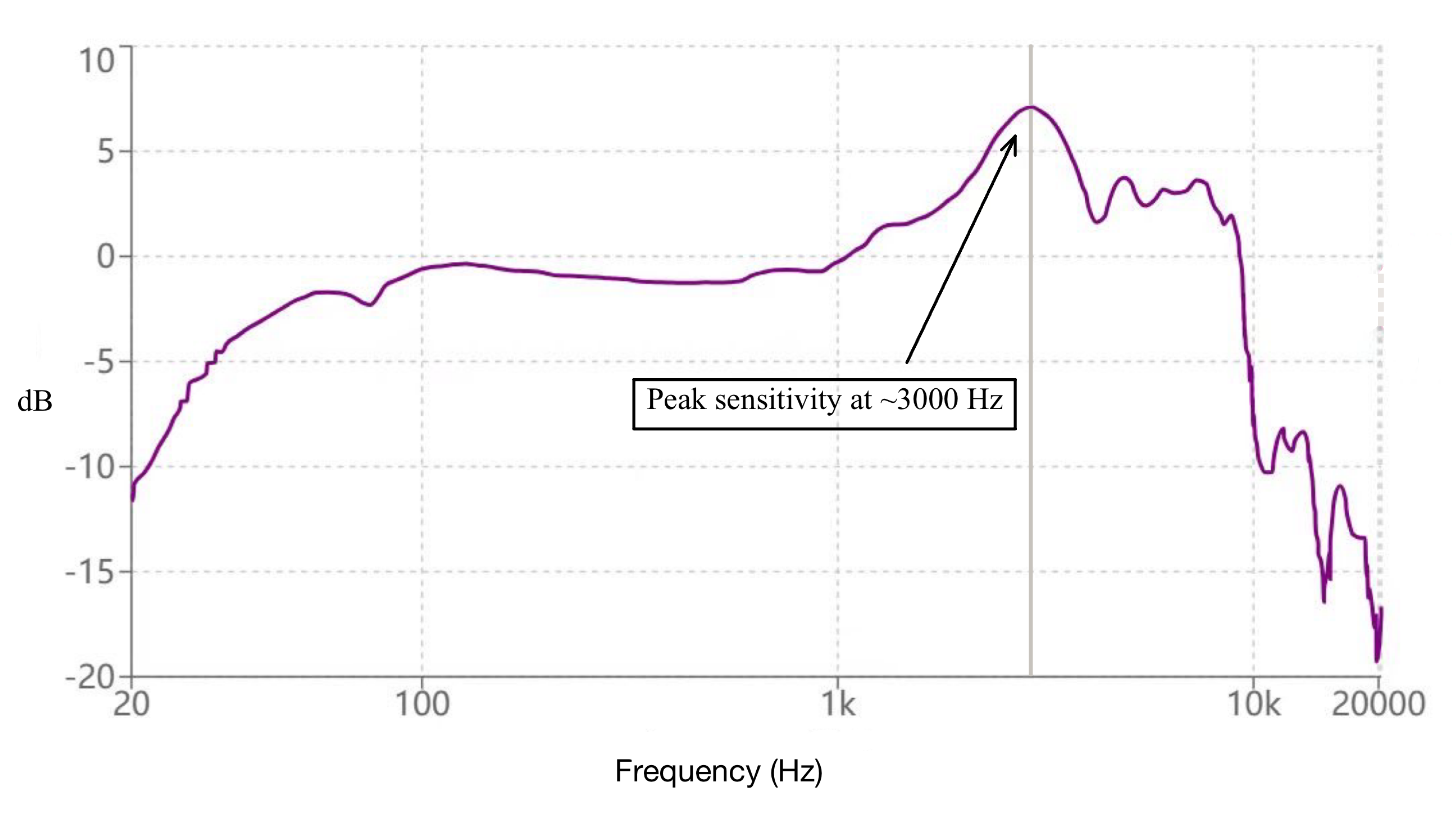}  
  \captionsetup{justification=centering}  
  \caption{Schematic diagram of human ear sensitivity to audio}
  \label{fig:image1}
\end{figure}
\subsection{Masking Threshold Calculation}

The masking effect, caused by the superposition of all critical bands on the $j$th critical band, is quantitatively estimated by convolving the energy of each critical band with the masking expansion function. The energy of each critical band, $B_j$, is calculated as follows. Let $X_j(i)$ represent the coefficient by which $X_c(k)$ and $X_d(k)$ are mapped to the $j$th critical band. The energy spectral density of the $j$th critical band is given by:
$ P_j(i) = X_j^2(i) $
where $j_l$ and $j_h$ represent the DCT coefficients corresponding to the lowest and highest frequencies of the $j$th critical band, respectively. The masking expansion function $SF_j(dB)$ is then computed based on this energy distribution\cite{Hu2016}.The constants inside are set based on experimental results and standard literature.
\begin{equation*}
\alpha_p = \frac{7.5}{\sqrt{17.5} \sqrt{1+(j + 0.474)^2}} \tag{1}
\end{equation*}
\begin{equation*}
SF_j = 15.81 + 7.5(j + 0.474) - \alpha_p \tag{2}
\end{equation*}
Calculate the extended masking threshold Cj.
\begin{equation*}
C_j = B_j * 10^{\frac{SF_i}{10}}\tag{3} 
\end{equation*}

In equation (3), * denotes a convolution operation. The magnitude of the masking threshold is influenced by the spectral flatness and tonal characteristics of the audio signal. The masking threshold for pure tone masking noise differs from that for noise masking a pure tone. In practice, signals are rarely pure tones or pure noise, so the masking threshold is corrected using a tonal coefficient. The masking threshold is quantitatively adjusted by estimating the spectral flatness and tonal coefficient of the $j$th critical band. The spectral flatness $SFM$ (in dB) of the critical band is then calculated as follows:
\begin{equation*}
SFM = 10\log\left(\frac{G}{A}\right)\tag{4}
\end{equation*}
In formula (4), G is the geometric mean of the critical band energy spectral density and A is the arithmetic mean of the critical band energy spectral density.
Calculate the pitch coefficient $\alpha$.

\begin{equation*}
\alpha = \min\left(\frac{SFM}{-60}, 1\right) \tag{5}
\end{equation*}
where $\alpha = 1$ when the audio signal is a pure tone and $\alpha = 0 $when the audio signal is white noise, and the pitch coefficient of the actual audio signal is often between [0, 1].
Calculate the masking threshold Tzj (dB).
The modified value of the jth critical band masking threshold obtained from the pitch coefficient is [1,5].
\begin{equation*}
O_j = a (14.5 + j)+(1 - a)5.5\tag{6}
\end{equation*}
The revised critical band j masking threshold is
\begin{equation*}
\tau_j = 10^{\lg C_j - O_j / 10} \tag{7}
\end{equation*}
It is considered that the estimated masking threshold correction value Tj should not be less than the absolute masking threshold Tq( f ) of the critical frequency band. The final masking threshold of the critical frequency band j is:
\begin{equation*}
\alpha_q = 6.5 e^{-0.6 \left( \frac{f}{1000} - 3.3 \right)^2}\tag{8}
\end{equation*}
\begin{equation*}
T_q(f) = 3.64 \left( \frac{f}{1000} \right)^{-0.8} - \alpha_q 
\quad + 0.001 \left( \frac{f}{1000} \right)^4\tag{9}
\end{equation*}Since $T_j$ represents the energy value and $T_q(f)$ denotes the sound intensity level, their units are not consistent, making a direct comparison of their magnitudes impractical. As defined by the sound intensity level and the absolute masking threshold, once the absolute energy spectral density corresponding to the 0 dB sound intensity level is determined, the absolute energy spectral density at each frequency point can be computed. In this paper, the absolute energy spectral density value corresponding to the 0 dB sound intensity level in the DCT domain is denoted as $S$. For computational convenience, the absolute masking threshold at the midpoint of each critical frequency band is taken as the absolute masking threshold $T_q(j)$ for that critical band. By substituting the center frequencies of the first 7 Bark domains into the calculation, $T_q(j)$ is obtained, and the corresponding absolute energy spectral density value is $S(j)$.

The final masking threshold for the critical band j is

\begin{equation*}
 T_z(j) = \max[T_j, I_j \times T_q(j)] = \max[T_j, I_j \times S(j)] \tag{10}
\end{equation*}
In formula (10), lj is the number of DCT coefficients in the jth critical band.
\subsection{Adaptive Strength Calculation for Pseudo-Timbre Embedding}

Based on the masking thresholds \(T_z(j)\) computed in the previous sections, we now determine the optimal embedding strength for the pseudo-timbre. This step is critical to ensure that the embedded pseudo-timbre remains imperceptible to human listeners while effectively interfering with voice conversion and text-to-speech models.

To achieve this balance, the audio signal is first decomposed using the Discrete Wavelet Transform (DWT) to obtain its time-frequency representation. With the masking thresholds for each critical band already established, the Quantization Index Modulation (QIM) technique is applied within each band to embed the pseudo-timbre. It is essential to control the quantization error introduced during this process to preserve the naturalness of the original audio.

First, for each critical band \(j\), the maximum quantization error \(\Delta_j\) is related to the corresponding noise energy by:
\begin{equation}
E_{n} = \Delta_j^2 \tag{11}
\end{equation}
Next, using the relationship between the noise energy and the masking threshold, the noise margin \(NMR(j)\) is determined as:
\begin{equation}
NMR(j) = E_{n} - T_z(j) \tag{12}
\end{equation}
where \(T_z(j)\) is the final masking threshold for the \(j\)th critical band, and \(NMR(j)\) represents the available margin for embedding. To ensure the embedding remains imperceptible, we enforce the condition:
\begin{equation}
10\log\left[NMR(j)\right] \leq -5 \tag{13}
\end{equation}
Finally, the adaptive embedding strength for the \(j\)th critical band is calculated by:
\begin{equation}
\Delta_j = \sqrt{\frac{10^{-1/2}\,T_z(j)}{I_j}} \tag{14}
\end{equation}
where \(I_j\) is a scaling factor that ensures consistency across bands.

By following these steps, we are able to embed pseudo-timbre into audio, preserving its naturalness and auditory intelligibility while effectively interfering with the training and generation process of AI VC models.

\section{EXPERIMENTAL RESULTS AND ANALYSIS}
\subsection{Experimental Setup}
\subsubsection{\textbf{Data Set}}
In this study, the CSTR VCTK and zhvoice corpus datasets are used for the experimental evaluations. The CSTR VCTK dataset \cite{yamaSgishi2019vctk} consists of recordings from 109 English-speaking speakers, each providing around 400 sentences. The zhvoice data set is a combination of eight open source datasets, which undergo noise reduction and silence removal. It contains approximately 3,200 speakers, 900 hours of audio, 1.13 million lines of text, and a total of about 13 million words. To ensure the universality and comprehensiveness of the experiment, the test data is divided into four categories: Chinese male voice, Chinese female voice, English male voice, and English female voice, with each category containing 100 sentences. The sentences in each category are selected using a stratified random sampling method to balance linguistic diversity and acoustic variability, thereby ensuring a fair and representative evaluation across all groups.

\subsubsection{\textbf{Defense Scenario}}

In the experiment, we aim to simulate real-world application scenarios by considering commercial and open-source clone speech models. Specifically, we tested the defense effectiveness against the commercial model ElevenLabs \cite{elevenlabs2023}, as well as the open-source models GPT-SoVITS \cite{gpt-sovits2024}, XTTSv2 \cite{coqui2023xtts}, SEED-VC \cite{wu2023seed}, and StyleTTS2 \cite{liu2023styletts2}. Additionally, in the comparative experiments, both Huang's and Wang's models employed AdaIN-VC as the pre-trained model for generating adversarial noise. The pseudo-timbre used in the models proposed in this study is available as open-source on GitHub.

\subsubsection{\textbf{Evaluation Metrics}}
In alignment with mainstream research, this experiment employs the speaker verification accuracy (SVA) metric to evaluate the defense effectiveness and perceptibility. SVA utilizes the ASV system \cite{snyder2018xvector} for speaker verification. The ASV system takes a pair of waveforms as input and returns a similarity score ranging from 0 to 1. If this score exceeds a predefined threshold of 0.8, the pair is deemed to come from the same speaker, thereby passing the ASV test. Conversely, if the score is 0.8 or below, the pair is considered to originate from different speakers. This threshold ensures a balance between precision and recall in the verification process, aligning with widely adopted practices in the field.

\begin{table}[ht]
\caption{Results after protection and Quality}
\centering
\resizebox{\linewidth}{!}{
\begin{tabular}{@{}c
>{\columncolor[HTML]{FFFFFF}}c 
>{\columncolor[HTML]{FFFFFF}}c 
>{\columncolor[HTML]{FFFFFF}}c 
>{\columncolor[HTML]{FFFFFF}}c 
>{\columncolor[HTML]{FFFFFF}}c 
>{\columncolor[HTML]{FFFFFF}}c c@{}}
\toprule
\cellcolor[HTML]{FFFFFF} &
  \cellcolor[HTML]{FFFFFF} &
  \multicolumn{5}{c}{\cellcolor[HTML]{FFFFFF}\textbf{Results after protection}} &
   \\ \cmidrule(lr){3-7}
\multirow{-2}{*}{\cellcolor[HTML]{FFFFFF}\textbf{Methods}} &
  \multirow{-2}{*}{\cellcolor[HTML]{FFFFFF}\textbf{Substitute models}} &
  \textbf{F\_Chinese} &
  \textbf{F\_English} &
  \textbf{M\_Chinese} &
  \textbf{M\_English} &
  \textbf{Average} &
  \multirow{-2}{*}{\textbf{Quality}} \\ \midrule
 &
  \textbf{ElevenLabs} &
  0.627 &
  0.442 &
  0.293 &
  0.371 &
  0.43325 &
  \cellcolor[HTML]{FFFFFF} \\
 &
  \textbf{GPT-SoVITS} &
  0.661 &
  0.465 &
  0.3 &
  0.508 &
  0.4835 &
  \cellcolor[HTML]{FFFFFF} \\
 &
  \textbf{XTTSv2} &
  0.553 &
  0.559 &
  0.326 &
  0.389 &
  0.45675 &
  \cellcolor[HTML]{FFFFFF} \\
 &
  \textbf{SEED-VC} &
  0.594 &
  0.531 &
  0.359 &
  0.596 &
  0.52 &
  \cellcolor[HTML]{FFFFFF} \\
\multirow{-5}{*}{\textbf{Ours}} &
  \textbf{StyleTTS2} &
  0.688 &
  0.73 &
  0.243 &
  0.238 &
  \textbf{0.47475} &
  \multirow{-5}{*}{\cellcolor[HTML]{FFFFFF}0.942} \\ \midrule
 &
  \textbf{ElevenLabs} &
  0.441 &
  0.534 &
  0.483 &
  0.304 &
  0.4405 &
  \cellcolor[HTML]{FFFFFF} \\
 &
  \textbf{GPT-SoVITS} &
  0.534 &
  0.433 &
  0.401 &
  0.528 &
  \textbf{0.474} &
  \cellcolor[HTML]{FFFFFF} \\
 &
  \textbf{XTTSv2} &
  0.623 &
  0.642 &
  0.495 &
  0.449 &
  0.55225 &
  \cellcolor[HTML]{FFFFFF} \\
 &
  \textbf{SEED-VC} &
  0.519 &
  0.49 &
  0.54 &
  0.519 &
  0.517 &
  \cellcolor[HTML]{FFFFFF} \\
\multirow{-5}{*}{\textbf{Huang's}} &
  \textbf{StyleTTS2} &
  0.414 &
  0.544 &
  0.533 &
  0.505 &
  0.499 &
  \multirow{-5}{*}{\cellcolor[HTML]{FFFFFF}\textbf{0.984}} \\ \midrule
 &
  \textbf{ElevenLabs} &
  0.515 &
  0.452 &
  0.304 &
  0.281 &
  \textbf{0.388} &
  \cellcolor[HTML]{FFFFFF} \\
 &
  \textbf{GPT-SoVITS} &
  0.61 &
  0.479 &
  0.411 &
  0.453 &
  0.48825 &
  \cellcolor[HTML]{FFFFFF} \\
 &
  \textbf{XTTSv2} &
  0.571 &
  0.428 &
  0.321 &
  0.424 &
  \textbf{0.436} &
  \cellcolor[HTML]{FFFFFF} \\
 &
  \textbf{SEED-VC} &
  0.515 &
  0.588 &
  0.524 &
  0.374 &
  \textbf{0.50025} &
  \cellcolor[HTML]{FFFFFF} \\
\multirow{-5}{*}{\textbf{Dong's}} &
  \textbf{StyleTTS2} &
  0.594 &
  0.417 &
  0.622 &
  0.314 &
  0.48675 &
  \multirow{-5}{*}{\cellcolor[HTML]{FFFFFF}0.956} \\ \bottomrule
\end{tabular}
}
\label{tab:1}
\end{table}

\subsection{Evaluation of Defense Effectiveness}
This section presents a comparison of three active defense methods:Huang's\cite{huang_defending_2021}, Dong's\cite{dong_active_2024} and Ours, evaluated using five models: ElevenLabs, GPT-SoVITS, XTTSv2, SEED-VC, and StyleTTS2. In the experiment, we use ASV (Automatic Speaker Verification) as the evaluation metric. ASV is an indicator that calculates the timbral similarity between the original audio and the cloned audio. A higher ASV value indicates greater similarity in timbre between the two audios. It is generally considered that when the ASV value exceeds 0.9, the two sounds are produced by the same individual.Table \ref{tab:1} illustrates the protective effectiveness of various active defense strategies against different cloned voices.  In this paper, a lower value of active defense indicates better performance of the active defense, while a higher quality value signifies better auditory effects after protection. The results presented in Table \ref{tab:1} reveal the following key findings:

(a) Significant Protective Effect:
1.The three methods exhibit varying degrees of protection across all clone speech models (including ElevenLabs, GPT-SoVITS, XTTSv2, SEED-VC, and StyleTTS2) on four different test datasets.
2.In particular, our active defense method (i.e., "Ours") demonstrates outstanding defensive capabilities against these clone speech models. In terms of ASV\_Similarity scores, the "Ours" method exhibits the strongest protective effect across multiple languages (both Chinese and English), indicating that our defense method excels in resisting adversarial attacks.

(b) Auditory Quality of Processed Models:
1.The models processed using our active defense method ("Ours") show exceptional auditory quality. The "Ours" method displays near-perfect speech quality in tests, particularly on F\_Chinese and F\_English.
2.Although other models exhibit higher similarity in certain cases (such as F\_Chinese and F\_English in the "Dong's" category), the speech quality of our method remains excellent, ensuring that the protected speech retains high similarity to the original speech while maintaining high intelligibility.
\begin{figure}[htbp]  
  \centering
  \includegraphics[width=\linewidth]{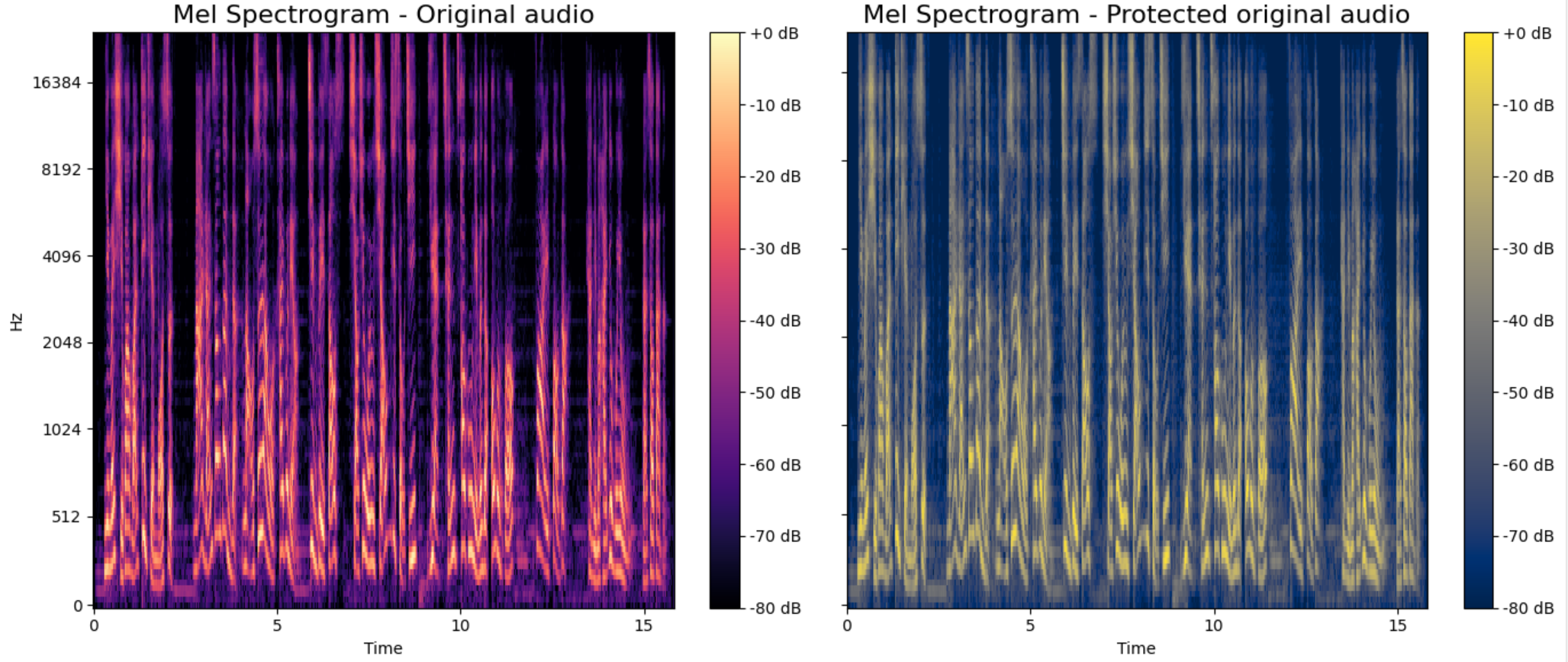}  
  \captionsetup{justification=centering}  
  \caption{Comparison of Mel spectrograms between original audio and protected original audio}
  \label{fig:4}
\end{figure}

These results suggest that our active defense method not only effectively defends against AI VC attacks but also strikes a remarkable balance between defensive efficacy and auditory quality.
\begin{table*}[ht]
\caption{Evaluation of Robustness}
\centering
\scriptsize
\begin{tabular}{@{}c
>{\columncolor[HTML]{FFFFFF}}c cccccccccc@{}}
\toprule
\cellcolor[HTML]{FFFFFF} &
  \cellcolor[HTML]{FFFFFF} &
  \multicolumn{10}{c}{\textbf{ASV\_Similarity↑}} \\ \cmidrule(l){3-12} 
\cellcolor[HTML]{FFFFFF} &
  \cellcolor[HTML]{FFFFFF} &
  \multicolumn{5}{c}{\cellcolor[HTML]{FFFFFF}\textbf{Results after noise reduction attack}} &
  \multicolumn{5}{c}{\cellcolor[HTML]{FFFFFF}\textbf{Results after compressed sampling rate}} \\ \cmidrule(l){3-12} 
\multirow{-3}{*}{\cellcolor[HTML]{FFFFFF}\textbf{Methods}} &
  \multirow{-3}{*}{\cellcolor[HTML]{FFFFFF}\textbf{Substitute models}} &
  \cellcolor[HTML]{FFFFFF}\textbf{F\_Chinese} &
  \cellcolor[HTML]{FFFFFF}\textbf{F\_English} &
  \cellcolor[HTML]{FFFFFF}\textbf{M\_Chinese} &
  \cellcolor[HTML]{FFFFFF}\textbf{M\_English} &
  \cellcolor[HTML]{FFFFFF}\textbf{Average} &
  \cellcolor[HTML]{FFFFFF}\textbf{F\_Chinese} &
  \cellcolor[HTML]{FFFFFF}\textbf{F\_English} &
  \cellcolor[HTML]{FFFFFF}\textbf{M\_Chinese} &
  \cellcolor[HTML]{FFFFFF}\textbf{M\_English} &
  \cellcolor[HTML]{FFFFFF}\textbf{Average} \\ \midrule
 &
  \textbf{ElevenLabs} &
  \cellcolor[HTML]{FFFFFF}0.579 &
  \cellcolor[HTML]{FFFFFF}0.391 &
  \cellcolor[HTML]{FFFFFF}0.403 &
  \cellcolor[HTML]{FFFFFF}0.442 &
  \cellcolor[HTML]{FFFFFF}0.45375 &
  \cellcolor[HTML]{FFFFFF}0.651 &
  \cellcolor[HTML]{FFFFFF}0.404 &
  \cellcolor[HTML]{FFFFFF}0.385 &
  \cellcolor[HTML]{FFFFFF}0.351 &
  \cellcolor[HTML]{FFFFFF}0.44775 \\
 &
  \textbf{GPT-SoVITS} &
  \cellcolor[HTML]{FFFFFF}0.512 &
  \cellcolor[HTML]{FFFFFF}0.436 &
  \cellcolor[HTML]{FFFFFF}0.295 &
  \cellcolor[HTML]{FFFFFF}0.176 &
  \cellcolor[HTML]{FFFFFF}\textbf{0.35475} &
  \cellcolor[HTML]{FFFFFF}0.445 &
  \cellcolor[HTML]{FFFFFF}0.457 &
  \cellcolor[HTML]{FFFFFF}0.361 &
  \cellcolor[HTML]{FFFFFF}0.203 &
  \cellcolor[HTML]{FFFFFF}\textbf{0.3665} \\
 &
  \textbf{XTTSv2} &
  \cellcolor[HTML]{FFFFFF}0.573 &
  \cellcolor[HTML]{FFFFFF}0.56 &
  \cellcolor[HTML]{FFFFFF}0.416 &
  \cellcolor[HTML]{FFFFFF}0.433 &
  \cellcolor[HTML]{FFFFFF}\textbf{0.4955} &
  \cellcolor[HTML]{FFFFFF}0.619 &
  \cellcolor[HTML]{FFFFFF}0.564 &
  \cellcolor[HTML]{FFFFFF}0.329 &
  \cellcolor[HTML]{FFFFFF}0.437 &
  \cellcolor[HTML]{FFFFFF}\textbf{0.48725} \\
 &
  \textbf{SEED-VC} &
  \cellcolor[HTML]{FFFFFF}0.532 &
  \cellcolor[HTML]{FFFFFF}0.58 &
  \cellcolor[HTML]{FFFFFF}0.302 &
  \cellcolor[HTML]{FFFFFF}0.661 &
  \cellcolor[HTML]{FFFFFF}\textbf{0.51875} &
  \cellcolor[HTML]{FFFFFF}0.551 &
  \cellcolor[HTML]{FFFFFF}0.566 &
  \cellcolor[HTML]{FFFFFF}0.391 &
  \cellcolor[HTML]{FFFFFF}0.670 &
  \cellcolor[HTML]{FFFFFF}0.5445 \\
\multirow{-5}{*}{\textbf{Ours}} &
  \textbf{StyleTTS2} &
  \cellcolor[HTML]{FFFFFF}0.476 &
  \cellcolor[HTML]{FFFFFF}0.396 &
  \cellcolor[HTML]{FFFFFF}0.338 &
  \cellcolor[HTML]{FFFFFF}0.63 &
  \cellcolor[HTML]{FFFFFF}\textbf{0.46} &
  \cellcolor[HTML]{FFFFFF}0.407 &
  \cellcolor[HTML]{FFFFFF}0.354 &
  \cellcolor[HTML]{FFFFFF}0.431 &
  \cellcolor[HTML]{FFFFFF}0.566 &
  \cellcolor[HTML]{FFFFFF}\textbf{0.4395} \\ \midrule
 &
  \textbf{ElevenLabs} &
  \cellcolor[HTML]{FFFFFF}0.672 &
  \cellcolor[HTML]{FFFFFF}0.839 &
  \cellcolor[HTML]{FFFFFF}0.625 &
  \cellcolor[HTML]{FFFFFF}0.801 &
  \cellcolor[HTML]{FFFFFF}0.73425 &
  \cellcolor[HTML]{FFFFFF}0.715 &
  \cellcolor[HTML]{FFFFFF}0.865 &
  \cellcolor[HTML]{FFFFFF}0.366 &
  \cellcolor[HTML]{FFFFFF}0.827 &
  \cellcolor[HTML]{FFFFFF}0.69325 \\
 &
  \textbf{GPT-SoVITS} &
  \cellcolor[HTML]{FFFFFF}0.672 &
  \cellcolor[HTML]{FFFFFF}0.783 &
  \cellcolor[HTML]{FFFFFF}0.524 &
  \cellcolor[HTML]{FFFFFF}0.745 &
  \cellcolor[HTML]{FFFFFF}0.681 &
  \cellcolor[HTML]{FFFFFF}0.595 &
  \cellcolor[HTML]{FFFFFF}0.758 &
  \cellcolor[HTML]{FFFFFF}0.443 &
  \cellcolor[HTML]{FFFFFF}0.751 &
  \cellcolor[HTML]{FFFFFF}0.63675 \\
 &
  \textbf{XTTSv2} &
  \cellcolor[HTML]{FFFFFF}0.742 &
  \cellcolor[HTML]{FFFFFF}0.72 &
  \cellcolor[HTML]{FFFFFF}0.695 &
  \cellcolor[HTML]{FFFFFF}0.648 &
  \cellcolor[HTML]{FFFFFF}0.70125 &
  \cellcolor[HTML]{FFFFFF}0.675 &
  \cellcolor[HTML]{FFFFFF}0.676 &
  \cellcolor[HTML]{FFFFFF}0.707 &
  \cellcolor[HTML]{FFFFFF}0.726 &
  \cellcolor[HTML]{FFFFFF}0.696 \\
 &
  \textbf{SEED-VC} &
  \cellcolor[HTML]{FFFFFF}0.448 &
  \cellcolor[HTML]{FFFFFF}0.618 &
  \cellcolor[HTML]{FFFFFF}0.763 &
  \cellcolor[HTML]{FFFFFF}0.704 &
  \cellcolor[HTML]{FFFFFF}0.63325 &
  \cellcolor[HTML]{FFFFFF}0.493 &
  \cellcolor[HTML]{FFFFFF}0.534 &
  \cellcolor[HTML]{FFFFFF}0.835 &
  \cellcolor[HTML]{FFFFFF}0.682 &
  \cellcolor[HTML]{FFFFFF}0.636 \\
\multirow{-5}{*}{\textbf{Huang's}} &
  \textbf{StyleTTS2} &
  \cellcolor[HTML]{FFFFFF}0.672 &
  \cellcolor[HTML]{FFFFFF}0.745 &
  \cellcolor[HTML]{FFFFFF}0.724 &
  \cellcolor[HTML]{FFFFFF}0.683 &
  \cellcolor[HTML]{FFFFFF}0.706 &
  \cellcolor[HTML]{FFFFFF}0.646 &
  \cellcolor[HTML]{FFFFFF}0.672 &
  \cellcolor[HTML]{FFFFFF}0.722 &
  \cellcolor[HTML]{FFFFFF}0.704 &
  \cellcolor[HTML]{FFFFFF}0.686 \\ \midrule
 &
  \textbf{ElevenLabs} &
  \cellcolor[HTML]{FFFFFF}0.565 &
  \cellcolor[HTML]{FFFFFF}0.431 &
  \cellcolor[HTML]{FFFFFF}0.424 &
  \cellcolor[HTML]{FFFFFF}0.374 &
  \cellcolor[HTML]{FFFFFF}\textbf{0.4485} &
  \cellcolor[HTML]{FFFFFF}0.503 &
  \cellcolor[HTML]{FFFFFF}0.402 &
  \cellcolor[HTML]{FFFFFF}0.420 &
  \cellcolor[HTML]{FFFFFF}0.358 &
  \cellcolor[HTML]{FFFFFF}\textbf{0.42075} \\
 &
  \textbf{GPT-SoVITS} &
  \cellcolor[HTML]{FFFFFF}0.394 &
  \cellcolor[HTML]{FFFFFF}0.413 &
  \cellcolor[HTML]{FFFFFF}0.325 &
  \cellcolor[HTML]{FFFFFF}0.313 &
  \cellcolor[HTML]{FFFFFF}0.36125 &
  \cellcolor[HTML]{FFFFFF}0.238 &
  \cellcolor[HTML]{FFFFFF}0.465 &
  \cellcolor[HTML]{FFFFFF}0.387 &
  \cellcolor[HTML]{FFFFFF}0.387 &
  \cellcolor[HTML]{FFFFFF}0.36925 \\
 &
  \textbf{XTTSv2} &
  \cellcolor[HTML]{FFFFFF}0.621 &
  \cellcolor[HTML]{FFFFFF}0.515 &
  \cellcolor[HTML]{FFFFFF}0.592 &
  \cellcolor[HTML]{FFFFFF}0.383 &
  \cellcolor[HTML]{FFFFFF}0.52775 &
  \cellcolor[HTML]{FFFFFF}0.619 &
  \cellcolor[HTML]{FFFFFF}0.576 &
  \cellcolor[HTML]{FFFFFF}0.566 &
  \cellcolor[HTML]{FFFFFF}0.345 &
  \cellcolor[HTML]{FFFFFF}0.5265 \\
 &
  \textbf{SEED-VC} &
  \cellcolor[HTML]{FFFFFF}0.594 &
  \cellcolor[HTML]{FFFFFF}0.429 &
  \cellcolor[HTML]{FFFFFF}0.519 &
  \cellcolor[HTML]{FFFFFF}0.578 &
  \cellcolor[HTML]{FFFFFF}0.53 &
  \cellcolor[HTML]{FFFFFF}0.552 &
  \cellcolor[HTML]{FFFFFF}0.503 &
  \cellcolor[HTML]{FFFFFF}0.437 &
  \cellcolor[HTML]{FFFFFF}0.574 &
  \cellcolor[HTML]{FFFFFF}\textbf{0.5165} \\
\multirow{-5}{*}{\textbf{Dong's}} &
  \textbf{StyleTTS2} &
  \cellcolor[HTML]{FFFFFF}0.521 &
  \cellcolor[HTML]{FFFFFF}0.454 &
  \cellcolor[HTML]{FFFFFF}0.396 &
  \cellcolor[HTML]{FFFFFF}0.486 &
  \cellcolor[HTML]{FFFFFF}0.46425 &
  \cellcolor[HTML]{FFFFFF}0.478 &
  \cellcolor[HTML]{FFFFFF}0.523 &
  \cellcolor[HTML]{FFFFFF}0.348 &
  \cellcolor[HTML]{FFFFFF}0.444 &
  \cellcolor[HTML]{FFFFFF}0.44825 \\ \bottomrule
\end{tabular}
\label{tag:4}
\end{table*}

\subsection{Evaluation of Robustness}
This section evaluates the robustness of three active defense methods. In practical applications, to better assess the protective effect on audio, attackers are typically categorized into two types: (a) amateur attackers and (b) professional attackers. Amateur attackers are generally assumed to use readily available online commercial products or simple pre-trained models. In contrast, professional attackers employ adversarial attacks, fine-tuning, and other advanced techniques to clone audio more effectively. For actively protected audio, these attackers might preprocess the audio using techniques such as denoising and downsampling.

In this chapter, we simulate the attacker's preprocessing steps, including denoising\cite{li2020frcrn} and resampling (resampling to 8 kHz), to test the robustness of the three methods. As shown in Table \ref{tag:4}, our method achieves an average score of 0.452 in the denoising attack experiment, significantly outperforming the other two methods. In the downsampling attack experiment, our method also achieves the highest average score of 0.46. In summary, our method has made a significant breakthrough in robustness, greatly enhancing its effectiveness in practical applications. We analyze the reasons for the poor robustness of traditional methods, noting that they all rely on GANs, which, although highly targeted in defense, are extremely vulnerable to destruction. As a result, their robustness and real-time performance are significantly inferior to our method.
\subsection{Generation Time Evaluation}

To maximize the protection of audio data, we aim to integrate active defense mechanisms into microphones or sound cards in practical applications. Therefore, real-time performance evaluations are also incorporated into our experiments. To assess the efficiency of our model in processing audio, we utilize Pyinstrument\cite{pyinstrument2023} to monitor the total processing time across the entire dataset. By dividing the total time by the number of audio samples in the dataset, we calculate the average processing time per audio sample.
\begin{figure}[htbp]  
  \centering
  \includegraphics[width=\linewidth]{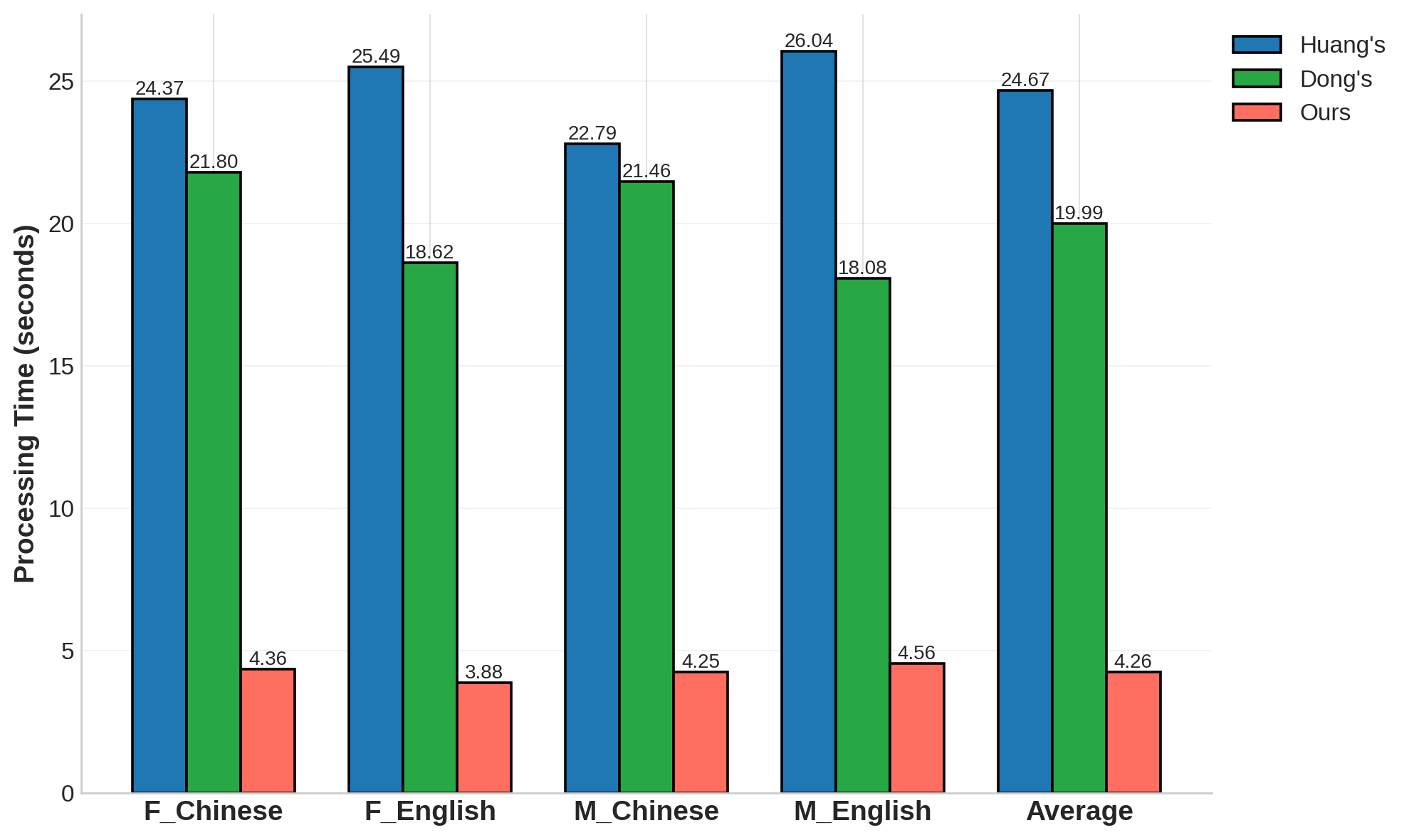}  
  \captionsetup{justification=centering}  
  \caption{Reciprocal of Processing Time Comparison of Different Models}
  \label{fig:3}
\end{figure}
As can be seen in Fig. \ref{fig:3}, our model has very superior performance, being 500\% faster than existing models.

\section{CONCLUSION}
This paper introduces a novel active defense mechanism, VocalCrypt, based on pseudo-timbre embedding using the masking effect, designed to counter AI-driven voice cloning attacks, specifically targeting speech conversion (VC) and text-to-speech (TTS) systems. We provide a comprehensive approach that utilizes the masking threshold principle to effectively embed imperceptible pseudo-timbres into audio signals, safeguarding speaker identities against unauthorized voice cloning attempts.

Our proposed method significantly outperforms existing defense strategies in both robustness and efficiency. Unlike traditional adversarial-based approaches that rely on post-detection or are vulnerable to noise reduction attacks, VocalCrypt proactively prevents voice cloning by manipulating the audio signal before it reaches the attacker’s system. This proactive intervention results in a highly effective defense mechanism, as demonstrated through extensive experiments using datasets from both English and Chinese speakers, including the Zhvoice and VCTK corpora. The results show that our approach is not only capable of preventing AI-generated voice forgery but also preserves the naturalness and intelligibility of the original voice, offering a balance between defense effectiveness and perceptual quality.

In terms of robustness, VocalCrypt demonstrates exceptional resistance to various attack types, including noise reduction and compression attacks. Furthermore, our method achieves remarkable real-time performance, outperforming existing methods by a factor of 500\% in terms of processing speed. This makes VocalCrypt a promising solution for real-world applications, where low-latency and high-efficiency defenses are critical.

Looking ahead, future work will focus on further enhancing the adaptability of VocalCrypt to different attack models, reducing the computational overhead even further, and improving its integration into hardware platforms for practical deployment in real-world scenarios. By continuing to refine the defense mechanism, we aim to provide an even more robust and scalable solution to protect against the growing threat of AI-driven voice manipulation.

\bibliographystyle{ieeetr}
\bibliography{main}

\begin{thebibliography}{10}
\bibitem{noauthor_artificial_nodate}
``Artificial intelligence makes voice cloning easy and 'the monster is already on the loose'.''
\end{thebibliography}
\end{document}